# The MEV Saga: Can Regulation Illuminate the Dark Forest?

Simona Ramos[1][0000-0003-0944-8891] and Joshua Ellul[2][0000-0002-4796-5665]

[1] Universitat Pompeu Fabra, Spain
[2] University of Malta, Malta

**Abstract.** In this article, we develop an interdisciplinary analysis of MEV which desires to merge the gap that exists between technical and legal research supporting policymakers in their regulatory decisions concerning blockchains, DeFi and associated risks. Consequently, this article is intended for both technical and legal audiences, and while we abstain from a detailed legal analysis, we aim to open a policy discussion regarding decentralized governance design at the block building layer - as the place where MEV occurs. Maximal Extractable Value or MEV has been one of the major concerns in blockchain designs as it creates a centralizing force which ultimately affects user's transactions. In this article, we dive into the technicality behind MEV, where we explain the concept behind the novel Proposal Builder Separation design (PBS) as an effort by Flashbots to increase decentralization through modularity. We underline potential vulnerability factors under the PBS design, which open space for MEV extracting adversarial strategies by inside participants. We discuss the shift of trust from validators to builders in PoS blockchains such as Ethereum, acknowledging the impact that the later ones may have on users' transactions (in terms of front-running) and censorship resistance (in terms of transaction inclusion). We recognize that under PBS, centralized (dominant) entities such as builders could potentially harm users by extracting MEV via front-running strategies. Finally, we suggest adequate design and policy measures which could potentially mitigate these negative effects while protecting blockchain users.

**Keywords:** MEV, Decentralized Design, Blockchain

## 1 Introduction

A basic rule of blockchains is that the most up-to-date state of the system is represented by the longest chain, and rational miners (under PoW) or validators (under PoS) are incentivized to attempt to generate new blocks that extend the chain further in order to gain the next block reward (and any fees associated with transactions included in the block). The higher the computational power of a miner (under PoW) or staking power of a validator (under PoS) the higher the probability of acquiring the ability to execute the next block is. Miners (or validators) receive transactions from various users and also may broadcast submitted transactions to other miners (or validators). Given the distributed and decentralized nature of the network, there is no way to know the order within which users' transactions were submitted — and therefore, it is up to the miners (or validators) to determine the order within which transactions are executed for the specific blocks a specific miner (or validator) is attempting to generate. Before transactions are included in a block, and after they are submitted into the network, they end up residing in a memory pool (mempool). Each miner (or validator) maintains its own mempool of pending transactions, and as discussed above, it is up to each individual miner (or validator) to decide on how to sort the transactions within the mempool. A node may decide to use a naive sorting strategy, where it simply appends transactions in the order the particular node received the transactions or it might perform a profit-maximizing strategies as discussed below.

The ability for miners (or validators) to order transactions turns them into a centralising force that creates an important challenge in the network which consequently impacts users. Miner (or maximum) extractable value (MEV) is a measure of the profit a miner (or validator, sequencer, etc.) can make through the ability to arbitrarily include, exclude, or re-order transactions within the blocks they produce — and, of course, miners (and validators) do their best to ensure they make maximum fees. As a result, this may lead to front-running, 'sandwiching' and other forms of market manipulation strategies, which can affect market prices, users' funds and the overall trust in the system. While MEV extraction can be considered an inherited part of the block building design, the evolution and adoption of Automated Market Makers (AMM) opened new arbitrage opportunities between crypto-asset markets bringing MEV to another level. Currently, statistics show that MEV is so pervasive that one out of 30 transactions is added by miners for this purpose [1], while sandwich attacks cost users more around 90 million dollars in 2022.

In traditional financial markets, user transactions are sequenced by a trusted and regulated intermediary in the order in which they are received. In a blockchain, by contrast, the updating of a block can be competitive and random. According to some critics, since these intermediaries can choose which transactions they add to the ledger and in which order, they can engage in activities that would be illegal in traditional markets such as front-running and sandwich trades, opening the discussion for the need of certain regulatory measures. There have been several open questions on whether current regulation on insider trading is directly transferable to MEV. Recently, the Bank of International settlements has emphasized this concern and asked for further regulatory research and adequate measures. From a general standpoint, MEV related challenges could be analyzed and potentially mitigated from two angles (one does not exclude the other):

1. by introducing technical solutions in the blockchain system and associated smart contracts, which includes developing incentive mechanisms under which the negative effects of MEV would be mitigated.
2. by introducing regulatory measures that could mitigate the negative impact of MEV and protect users and other affected parties.

Overall, in order to make effective policy, a solid understanding of the technicality behind MEV is needed as regulatory solutions can find it useful to follow the current development on the technical side - in order to understand where things could go wrong (e.g., where centralization may occur).

## 2   MEV basics: The dark forest & Flash Boys 2.0

The ability of miners to access the mempool and rearrange transactions in accordance to perceived fee value has been at the core of MEV. In general, there are harmful and unharmful activities that involve MEV. For example, arbitrage and liquidations are noted as potentially bening activities (unharmful) which tend to promote market efficiency [2]. On the other hand, the harmful ones have predominated in market discussions as they have dramatically increased in the past few years costing users millions of dollars.

While some articles regarding MEV focus on miners as profit maximizing entities that utilize mempool information to generate extra profits [1], the current system design enables other adversarial entities to target user transactions by creating diverse types of attacks. For example, currently many users have suffered from adversarial actions (such as front-running, back running, sandwich attacks, etc.) done by very specialized 'Arbitrage Bots' that detect arbitrage opportunities across the network and replicate users transactions with a higher gas price hence managing to extract additional value, and overburden the system by creating bot-to-bot competition attacks. Arbitrage bots constantly monitor pending transactions in the mempool and are able to rapidly detect and exploit profitable opportunities.

As emphasized in one of the earliest articles on MEV- *"Ethereum is a Dark Forest"*, the authors explain a situation where front-running arises because the transaction broadcasted by the legitimate claimant to a smart contractual payment, can be seen and slightly altered by others — specifically, arbitrage bots — to direct token payment to an alternative adversarial owned wallet [3]. By offering higher transaction fees and leveraging on the lag involved in this process, these bots can have the same transaction recorded with an earlier time-stamp, on the same or on an earlier executed block than the legitimate claimant, hence making the bot's transaction valid and overruling the transaction of the legitimate claimer. This front-running example is very similar to the one described in [4]. There, front-running involves racing to take advantage of arbitrage opportunities that are created in the nanoseconds after someone engages in an asset purchase on financial markets but before the transaction has reached the market. Likewise, this type of front-running often occurs when bots try to take over arbitrage opportunities between cryptocurrency exchanges.

Alongside front-running, the most common MEV attacks also include back running, and sandwich attacks. Sandwich attacks are common adversarial technique. For a sandwich attack to occur, imagine that Josh wants to buy a Token X on a Decentralised Exchange (DEX) that uses an automated market maker (AMM) model. An adversary which sees Josh's transaction can create two of its own transactions which it inserts before and after Josh's transaction. The adversary's first transaction buys Token X, which pushes up the price for Josh's transaction, and then the third transaction is the adversary's transaction to sell Token X (now at a higher price) at a profit. Since 2020, total MEV has amounted to an estimated USD 550–650 million just on the Ethereum network [1]. MEV can also essentially increase the slippage in the trading price for users. Slippage is a de facto a 'hidden price impact' that users experience when trading against an automated market maker (AMM). When trading via an AMM, the expected execution price can differ from the real execution price because the expected price depends on a past blockchain state, which may change between the transaction creation and its execution because of certain actions (e.g., front-running transactions).

## 3     PBS: How does it work and why is it important

Proposer/Builder Separation (PBS) is a blockchain design feature that divides block building into the roles of block proposers and block builders. Block proposal is the action of submitting a block of transactions for the approval of network validators, while block building is the action of transaction ordering. When a blockchain protocol separates these two actions, it simplifies the process of completing each task and allows actors to specialize in one or the other. On most blockchains, a singular actor completes both tasks. For example, before Ethereum completed 'The Merge' there was no pro-proposer/builder separation and miners had a sole control.  Arguably, proposer/builder separation (PBS) mitigates these problems by splitting the block construction role from the block proposal role. In simplest terms, at first, users/searchers send transactions to block builders through public or private peer-to-peer transaction pools. A separate class of actors called **builders** are responsible for building the block bodies—essentially an ordered list of transactions that becomes the main "payload" of the block, and submit bids. **Block proposers** receive a block from their local block builder, and sign and propose it to the network. For their work, the chosen builder receives a fee from the validator after the execution of the block [5].

An important party in the block building ecosystem, Flashbots, play a crucial role in this system [5]. Considered a type of entity providing a public good, Flashbots focus on mitigating the existential risks MEV could cause to stateful blockchains like Ethereum. In essence, Flashbots provides a private communication channel between Ethereum users and validators for efficiently communicating preferred transaction order within a block.  Flashbots connect users/searchers to validators while allowing them to avoid the public mempool.

As noted in Figure 1. below, searchers (users) may send transactions via so called bundles through a block builder such as Flashbots itself.  Bundles are one or more transactions that are grouped together and executed in the order they are provided. The builder simulates the bundles to ensure validity of transactions, and then builds a full block. In this way users can 'hide' their transactions (avoid public mempool) before they are publicly executed in a block. Hence, users/searchers can avoid potential front-running and other types of adversarial attacks by using Flashbots.

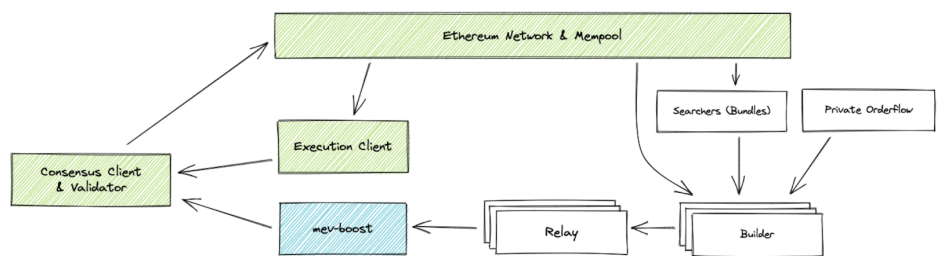

**Fig. 1.** PBS Design Scheme. Source: flashbots.net

Whilst PBS is optional for validators, as they can decide to arrange transactions in their own way and extract MEV, PBS is beneficial for them as it minimizes validator computational overhead. Hence it is likely that rational validators would eventually resort to using PBS. However, PBS ultimately incentivizes builder centralization, shifting the need for validator trust to builder trust. In essence, PBS does not fully avoid front-running attacks, as these can still be done by builders. Builders can still act as searchers and include their own MEV extracting transaction, ultimately front-running the user's delegated transaction.

In the effort to increase modularity and increase democratization amongst builders, Flashbots also introduced MEV-boost as part of the PBS structure. MEV-boost is open source middleware which helps create a competitive block-building market. MEV-boost act as a trusted party and aggregate blocks from multiple builders and identify the most profitable block to submit to the block proposer (validators). At the moment this part is far from truly decentralized. Data from mevboost.org shows that there are six major active relays currently delivering blocks in Ethereum, namely Flashbots, BloXroute Max Profit, BloXroute Ethical, BloXroute Regulated, Blocknative and Eden.

There are still certain risks that can occur from PBS. Identified risks include: *Builder Centralization; Builder/Relay Collusion and Malicious Relays*. Understanding the risks associated with centralization and collusion are important as they open a regulatory discussion as we explain in the next section. Arguably, a centralized builder/relay ecosystem, gains the ability for censorship and access to exclusive transaction order flow from which front-running types of attacks can be executed. Ultimately this creates market inefficiencies and impacts users negatively. For example, consider a wallet trying to send exclusive order flow to a single builder. For this order flow to be executed, it has to be included in a block on the blockchain which may take time. In order to avoid execution delays, a rational user will minimize this delay by sending the order to the builder with the highest inclusion rate, further increasing their dominance and centralization of the market. In that case, exclusive order flow would allow a builder, or small group of colluding builders, to capture the builder market, making it effectively uncompetitive. Also, a dominant builder would have a significant amount of private transaction information, allowing him to be in a more privileged position to extract MEV through front-running.

## 4     Can Regulation illuminate the Dark Forest?

As demonstrated in the paragraphs above, MEV can negatively affect market participants — and to a large extent. Arguably MEV is still an under researched topic and has been shown to be an evolving force (as demonstrated through proposed changes for block building). Besides certain theoretically derived scenarios noted by [6], there are no other studies (to the best of the authors' knowledge) that investigate the impact of MEV on consumers/users/investors — an important factor that could merit regulatory intervention. Hence, we have resorted to looking at MEV from a security perspective - trying to understand when and how things could go wrong for users, in terms of attacks and vulnerabilities exploits (such as sandwiched user transactions) or in terms of transaction censorship (such as purposely excluding user transactions from the building block). We agree with [2] argument on the need for a more well-defined and detailed differentiation of the notion of what a "victim" is in an MEV scenario. In other words, someone that has been deprived of something that was rightfully theirs may not be the same as someone whose 'transaction or trade-based profit' decreased due to certain market movement. Arguably, regulation is not meant to protect users by enabling conditions for them to achieve profit maximizing. Also, this definition may be interpreted differently as non-adversarial forms of MEV extractions (e.g., liquidation) may still indirectly render a negative impact on users via its effect on the market as a whole. In this paper, we follow [7] and [2] differentiation of MEV under three broad categories: Monarch, Mafia and Moloch.

- "Monarch" extractable value refers to the more broadly accepted understanding of MEV, as value extractable due to the power to order and allocate (block) space.
- "Mafia" extractable value "arises when one agent (coalition of agents) gains an asymmetric knowledge of another agent's private information (asymmetric sophistication)."
- "Moloch" extractable value arises from inefficient coordination methods.

In the following sections, unlike [2] we focus mostly on Monarchs (which can extract value based on their ability to dominate transaction ordering) and Mafia — referring to value that arises when an entity has asymmetric information of users' transaction information. We look at these two occurrences as potentially being performed by 'insiders' such as builders due to their ability to create adversarial and market manipulative strategies in order to achieve profit maximizing.

### 4.1 Front-running in Traditional Finance vs. DeFi

In traditional finance, front-running is considered unethical and often illegal. The premise of illegality is based on the notion that a trader possesses and acts on inside information to achieve personal gains. The non-public information concerning certain transactions ought to be of a 'material size' - meaning that it is significant enough to cause a price change in the futures or options contract and thereby allow the front-runner to profit [8]. An example of front-running in traditional finance is when a broker exploits significant market knowledge that has not yet been made public. This is similar to insider trading, with the minor difference that the broker works for the client's brokerage rather than inside the client's business [9]. In reality, front-runners profit by exploiting the discrepancy between the security's true value and its market value [8]. Consequently, the market price of the security being bought and sold does not reflect the true value of the security which distorts market efficiency. Enforcement of insider-trading regulations is currently a high priority for the Securities and Exchange Commission (SEC) [10].

The possibility that insider information has impacted cryptocurrency returns has been noted in [11], where the authors show the impact of cyber-attacks related news on price changes of the attacked cryptocurrency (in terms of abnormal losses). This example however mostly involves the centralized user-facing layer in the blockchain ecosystem (e.g., users trading via centralized exchange) where insider information can mostly be used by outsiders (e.g., hackers). Nevertheless, front-running at the block-building layer can be exploited by insiders - due to the ability of miners to arbitrarily include, exclude, or reorder transactions in blocks, allowing them to even place their own transactions when they identify a profitable opportunity. This can be seen to be a much more serious case of front-running as it is not due to abnormal temporal shocks to the system (such as an unexpected cyber-attack) but as a byproduct of the block-building design.

In essence, miners can be noted as possessing insider information, however this is due to the technical constraints that create a lag in the recording of transactions across all nodes in different geographical locations. For example, different nodes (e.g., a node in Australia and a node in Alaska) would not have a same view of the mempool's list of transactions, although there would be some overlap and similarities. This happens because every node receives transactions from different neighbouring nodes at different times.

When it comes to transaction ordering under the PBS designs, users send transactions to builders via a private channel. This gives builders the ability to have insider information over potential front-running opportunities. As noted in [5], a centralized builder ecosystem of a builder that dominates the market because of its outsized profitability gains the ability for censorship and access to exclusive transaction order flow. According to [2] under the PBS design, a user may send a transaction only to one builder, placing this entity in a privileged position and enabling them to treat user transactions as a kind of private order flow. The way the builder may arrange transactions could result in front-running or other types of adversarial attack which may impact the user's transaction gains negatively. Nevertheless, in this case there would be a loss of trust to the selected builder party which could disincentivize the user to send private flow transactions via this entity again.

## 4.2 Regulating central points in a decentralized world

Nissembaum approaches trust and security in ICT as a conglomeration of two main factors, namely composed of insiders (e.g. miners, validators, builders, etc.) and outsiders (e.g. hackers), maintaining that very often issues can appear from an `adversarial insider' behavior [12]. This perspective has been essential when analyzing blockchains, as the system operations are primarily maintained by 'insiders', whose dishonest or adversarial behavior may put the security and reliability of the system at stake [13].

The incentive design behind a blockchain system is crucial for its proper operation, however under certain premises of some blockchains networks, participants can engage in strategic behavior by exploiting profitable opportunities (e.g., MEV). On one hand, some non-malicious forms of `collusive behavior', such as miners joining mining pools, may be essential for miners to beneficially participate in the system (e.g. due to the increased difficulty level of generating a block), although under certain premises mining pools can collude and act as a cartel and threaten the trustworthiness of the system. [14] argues that on-chain conduct may render issues of tortious liability for miners. However, treating miners as fiduciaries could discourage them from participating in what may be considered a socially beneficial project, due to a fear of potential liability, and without them contributing processing power the system risks disappearing.

Blockchain based ecosystem there might be little effectiveness of regulation of a system that is composed of a decentralized and distributed (also no-easily-identifiable) group of entities [15] [16]. Indeed, under the PoW design, regulating miners would be difficult due to their anonymity/pseudonymity trait. While a front-running activity can be noted on the chain, the pseudo-anonymous trait of miners makes it very difficult to regulate. As such, deterrence based on sanctioning in a decentralized setting may not be highly effective. According to [13], sanctioning measures are less likely to be effective in the cyber sphere where the identity of the attacker is uncertain and there are many unknown adversaries.

However, we note that whilst regulating decentralized networks of many different miners can be difficult, regulating small networks can be more easily achieved — since with enough investigative effort it may be possible to identify the few miners involved (e.g. pseudonymous illicit actors are often identified via some off-chain interaction). Indeed, such a small network would be decentralized only to those participants and outsiders would likely deem the network as 'centralized' — yet it is important to highlight that de/centralization is not a binary value, but is a spectrum (from a system being highly decentralized to less and completely centralized when involving a single actor). Arguably, blockchain is a more complex ecosystem where different liability rules may apply depending on a careful distinction of the modularity of the layer and the given use-case. For example, under a PBS design the anonymity trait of builders is reduced.

This is because, while anyone can be a miner, builder entities tend to be more organized due to the overhead needed in their operation.

Moreover, the lack of any single point of failure, dominance and control contributes to increasing the resilience of blockchain-based systems, while also making it more difficult for national laws to be enforced on them. Nevertheless, in practice, this might not be the case, as centralization has occurred in various segments across the blockchain ecosystem, including the block building layer. However, a propensity for centralization does not necessarily mean that the dominant entity would exert its power to over-run the system. In other words, while centralization is an undesired occurrence in decentralized systems, it doesn't necessarily mean that would negatively affect users. However, this situation may ultimately reduce the trust in the underlying ecosystem and disincentivize participation - which puts the system at a risk for survival.

Centralization of builders through collusion can happen under the PBS design which can allow builders to extract large amounts of MEV. As we note the previous section, an exclusive overflow leads to builder centralization, creating dead-weight loss on the user side as a consequence of the market inefficiencies. Furthermore, under the PBS design, there is possibility for builder collusion which could lead to increased interblock front-running due to profitable opportunities related to transaction ordering [5].

Overall, there are different types of front-running. In our case, there are two relevant categories: intrablock (frontrunning happens in the same block) and interblock (multi-block front-running). The former can be done (in general) with or without builder collusion. The second, on the other hand, is difficult to execute without collusion. Overall, if the builders' collusion seeks to maximize profits then the second type of front-runnings would occur, which would imply that the first would also occur. This creates a need to avoid collusion among builders, for which regulation can prove to be useful under certain conditions. Consequently if, a) adversarial builder collusion can be detected and b) regulation has means to detect and punish builders – than applying certain type of monetary sanctions may render useful in deterring malicious behavior (e.g., front-running). For example, in a simplistic game of two builders colluding (where payoff is higher for both than not colluding), and if there is a regulatory penalty for collusion when detected, then depending on the size of the front-running profit and size of penalty, builders may be disincentivized to collude and front-running will be reduced. In other words, a suitable size of penalty ought to be established depending on front-running profits for which builders would not be willing to risk and collude.

Another measure to reduce collusion is for governments or other trusted institutions to have their own builders, however this would imply a partial switch of trust in such trusted institutions and a potential for increase in censorship. Government interference at block building level has shown in practice to have negative impact on censorship resistance in blockchains. As noted, after OFAC (Office of Foreign Assets Control in the USA) sanctioned Tornado Cash and several Ethereum addresses associated with it, there was a noted difference in block building (transaction inclusion) – as blocks stopped including transactions coming from the mixer (non-OFAC compliant) [17]. This basically means that base layer participants would either exclude any sanctioned addresses in blocks they propose or refuse to attest to any blocks that include such sanctioned addresses. Another important insight is the presence of MEV-Boost in the blocks. Interestingly, most of the blocks involve a MEV-boost relay which is an essential part of the overall selection (and execution) of blocks that involves transaction ordering [18]. Arguably, sanctioning mixers does have an impact on overall transaction inclusion and block creation, ultimately affecting the decisions by the still heavily centralized block builders and relays (MEV-boosts) under the PBS proposed design. Also, imposing fiduciary duties may also render negative effects as a regulatory arbitrage may appear when operators may decide to register in a place of favorable jurisdiction while being able to operate in another [2].

Alternatively, a type of monitoring measure that can be used by governments to make sure that transaction ordering is random (hence no strategy for front-running) is to oblige builders to use TEE (Trusted Execution Environment). Such TEE code could be programmed and then audited by governments to assure transaction ordering takes place in the desirable manner (e.g., random). While this would in theory avoid front-running by builders, it would increase the overall cost of using the system, as it adds another step in the process (increasing fees) and would decrease overall welfare creation. Moreover, this could potentially incentivize spamming the builder, as a malicious player could send many transactions which would increase the likelihood of one of the transactions being placed in front of the targeted transaction. Overall, it is important to mention that regulatory measures placed at the block building layer in blockchains amounts to an 'inability' to report a transaction due to a fear of liability; not an ability to "fully block" it as it is usually done in traditional finance.

Arguably, trying to allocate 'potentially central points' in order to apply centralized (traditional) regulatory measures may show to be partially effective as it may undermine some of the most important traits of the system (such as censorship resistance). As maintained by [19] and cited by [20]: *"Trying to apply centralized solutions to decentralized problems fails. It fails to scale, and it fails to achieve any of the stated goals. Although, it does push the decentralized platforms to try to innovate elsewhere ... The answer is really simple. If you want to solve decentralized problems, solve them with decentralized solutions.... "*

An alternative with merits further discussion is the ability of the system for self-regulation. In line with Lessing's "code is law", an argument can be made for the ability for on-chain governance to incorporate certain constitutional (regulation-like) principles into its block-building design which would imply a certain level of regulation and protection from colluding and centralized powers. For example, as a desired outcome of the PBS design - if in the future a certain optimal is achieved as to create a competitive market for builders - a reputation system can be established as a form of self-regulation and user protection. In other words, a reputation system can be created where users could record if they suspect that any of their transactions were front-runned or sandwiched. This could serve as a security assurance for users, when selecting which builder to send their transaction to. Consequently, this type of system would reinforce trust in a potentially centralized setting. Likewise, on-chain incentive design suggestions imply a creation of a 'fee escalator system.' Under this system market inefficiencies seem to be corrected (in theory) as the user is put in a seemingly powerful position to run an auction "facing the other way" in the MEV supply chain. In other words, the situation can be flipped so that MEV extractors offer bids to users to execute their order. In line with [2], on-chain based self-regulation would avoid the high cost of regulatory intervention, such as monitoring, detection and enforcement of penalties.

## 5     Conclusion

In this article we dived into the technology underpinning MEV, shining light on the evolution of the block building design at the consensus layer. We explained the novel Proposal Builder Separation (PBS) design and underlined potential vulnerability factors associated with it - such as the potential for builder centralization and collusion in order to front run user's transactions and maximize profits. While we recognize that under PBS, centralized (dominant) entities such as builders could damage users (and the system) by extracting MEV via front-running types of attacks, it is a matter of a) technological detection techniques advancements and b) regulatory ability to identify and punish - that regulatory mechanisms could be established to protect users. Nevertheless, we warm over the potential negative effects of regulatory intervention due to their impact on censorship resistance. Likewise, taking into consideration the cost of regulatory supervision and enforcement, technical (detection-based techniques) measures may suffice. In other words, detection of malicious activities by builder would ultimately reduce the trust and confidence in this entity, reducing the number of users transacting through that network. Overall, this might have a short run negative impact on the network (and the amount of transactions being executed) until confidence is restored and certain safeguarding principles are established. Moreover, alluding to one of the main EU's objectives regarding new technology law: "to protect users, without withholding innovation", we argue that regulating at the base layer in public blockchains may disincentivize main participants to contribute effort (e.g., staking, computing power, code, equipment etc.) and thus pose a risk for the system to further centralize or disappear. Overall, on-chain measures such as reputation design and fee escalator system are suggested as viable and cost-effective.